\documentclass[12pt]{iopart}
\usepackage{graphicx}
\usepackage[export]{adjustbox}

\begin{document}

\title[Experimentally optimizing QKD rates via nonlocal dispersion compensation]{Experimentally optimizing QKD rates via nonlocal dispersion compensation}

\author{Sebastian Philipp Neumann$^{1,2}$*, Domenico Ribezzo$^{1,2}$, Martin Bohmann$^{1,2}$ and Rupert Ursin$^{1,2}$*}

\address{$^1$ Institute for Quantum Optics and Quantum Information Vienna, Austrian Academy of Sciences, Boltzmanngasse 3, 1090 Vienna, Austria}
\address{$^2$ Vienna Center for Quantum Science and Technology, Boltzmanngasse 5, 1090 Vienna, Austria}
\ead{sebastian.neumann@oeaw.ac.at, rupert.ursin@oeaw.ac.at}
\vspace{10pt}
\begin{indented}
\item[]January 2021
\end{indented}

\begin{abstract}
Quantum key distribution (QKD) enables unconditionally secure communication guaranteed by the laws of physics.
The last decades have seen tremendous efforts in making this technology feasible under real-life conditions, with implementations bridging ever longer distances and creating ever higher secure key rates.
Readily deployed glass fiber connections are a natural choice for distributing the single photons necessary for QKD both in intra- and intercity links.
Any fiber-based implementation however experiences chromatic dispersion which deteriorates temporal detection precision.
This ultimately limits maximum distance and achievable key rate of such QKD systems.
In this work, we address this limitation to both maximum distance and key rate and present an effective and easy-to-implement method to overcome chromatic dispersion effects.
By exploiting entangled photons' frequency correlations, we make use of nonlocal dispersion compensation to improve the photons' temporal correlations.
Our experiment is the first implementation utilizing the inherently quantum-mechanical effect of nonlocal dispersion compensation for QKD in this way.
We experimentally show an increase in key rate from 6.1 to 228.3\,bits/s over 6.46\,km of telecom fiber.
Our approach is extendable to arbitrary fiber lengths and dispersion values, resulting in substantially increased key rates and even enabling QKD in the first place where strong dispersion would otherwise frustrate key extraction at all.
\end{abstract}

%
\vspace{2pc}

\section{Introduction}
Quantum key distribution (QKD) enables communication partners to exchange messages with unconditional cryptographic security based on the laws of quantum physics rather than assumptions about computational hardness.
This decisive advantage of QKD over classical encryption techniques has stimulated intensive research since its first proposal in 1984~\cite{Bennett2014}.
The key challenge in state-of-the-art QKD research is the development of feasible strategies enabling the faithful distribution of quantum states of light over long distances and at high rates~\cite{Xu2020}.
Glass fibers are an obvious choice for quantum communication, since existing telecommunication infrastructure can be used and links can be operated 24/7 independent of weather conditions, in contrast to satellite connections~\cite{Yin2020}.
Fiber links are versatile. They can be deployed in network configurations for intra-city links over tens of kilometers~\cite{Joshi2019, Dynes2019, Bacco2019} and have also been used for long-distance communication under laboratory conditions~\cite{Korzh2015,Yin2016,Boaron2018a}.
Only recently, in-field QKD using commercially deployed fibers has been shown over a 96\,km submarine link~\cite{Wengerowsky2019}. Another in-field connection over 66\,km withstood noise from classical traffic in another wavelength channel of the same fiber~\cite{Mao2018}.

The performance of a QKD system is quantified by its secure key rate.
Naturally, this rate can be enhanced by sending more photons per time unit, but such an increase in photon creation rate is only beneficial as long as successively sent photons can still be unambiguously identified in time by both communication partners.
If identification cannot be guaranteed with sufficient fidelity, the existence of an eavesdropper in the quantum channel cannot be ruled out anymore and no secure key is created.
Chromatic dispersion in fiber however induces substantial temporal overlaps of single photons.
This is because it causes photons at different parts of the photon source’s wavelength spectrum to travel at different speeds in the fiber.
Dispersion therefore forces experimenters to dim their single photon source to levels far below what would be technically possible~\cite{Steinlechner2013, Meyer-Scott2018} in order to tell consecutively sent photons apart.
This poses a substantial limit on the secure key rate of any fiber-based QKD protocol.

Assuming state-of-the-art detectors, this effect comes into play already for 10\,km fiber links conforming to the International Telecommunication Union's (ITU) most popular fiber and multiplexing standards~\cite{ITUG652, ITUG694, Warier2018}, which we will use for all further calculations if not noted otherwise.
For longer distances, the effect becomes ever more extensive, until key production is completely prevented at around 460\,km fiber length, assuming etangled photons with 100\,GHz spectral width (see \ref{chap:limits}).
Importantly, dispersion negatively affects any QKD protocol, no matter which degree of freedom is carrying the quantum information and which particular protocol is used.
This is also true for high-dimensional quantum information protocols \cite{Cozzolino2019} where more than one qubit per photon is transmitted.
Therefore, dispersion poses a key challenge in fiber-based quantum communication, and has to be addressed by any future implementation.

Entanglement-based QKD protocols such as BBM92~\cite{Bennett1992} or device-independent protocols \cite{Acin2007} offer a unique method of overcoming dispersion-induced performance degradation.
An entangled photon pair's correlations in time and energy allow for so-called nonlocal dispersion compensation~\cite{Franson1992}.
Such a compensation scheme uses the photons' (anti-)correlations in wavelength as a resource to tighten temporal correlations which have been dissolved by chromatic dispersion.
Correlations in polarization, which are used for key creation in this work, are left intact in this process.

In this work, we experimentally overcome the detrimental effect of chromatic dispersion on QKD rates for the first time.
In particular, we integrate a nonlocal dispersion compensation scheme into a full-fledged polarization-based BBM92 protocol over 6.46\,km of telecom fiber.
Canceling dispersion in this way, we enable polarization-state measurements with low error rate, thus significantly increasing the implementation's performance.
In a realistic scenario of high loss, we report an increase of the secure key rate from $6.1$ to $228.3$\,bits/s, i.e. by a factor of $37$, with the method described.

\section{Methods}
\subsection{Mitigating dispersion effects}
The total timing uncertainty a QKD protocol is subjected to can be written as
\begin{equation}
   \Delta T=\sqrt{\sigma_\mathrm{C}^2+\sigma_\mathrm{J}^2+\sigma_\mathrm{D}^2}.
\label{eq:deltaT} 
\end{equation}
Here, $\sigma_\mathrm{C}$ is the photons' coherence time, which is a fundamental property related to their finite spectral width~\cite{SalehTeich}. $\sigma_\mathrm{J}$ signifies timing jitter due to imperfect detection electronics, and $\sigma_\mathrm{D}$ is the temporal spread due to chromatic dispersion. We assume independent normal distributions for each effect.

For photons of $100$\,GHz ($\approx 0.8$\,nm) spectral width at $1550$\,nm, the coherence time $\sigma_{\mathrm{C}}$ is less than $5$\,ps.
This is negligible for current BBM92 applications.
Regarding timing jitter $\sigma_{\mathrm{J}}$, superconducting single-photon nanowire detectors (SSPD) are the state of the art.
The lowest SSPD jitter values reported today are in the order of $5$\,ps for telecom wavelengths~\cite{Korzh2020}, while commercial devices including time-tagging electronics typically exhibit jitters of $40$\,ps~\cite{SQ2020, IDQ, Swabian}.
The steady advancements in nanowire technology, however impressive they may be, can nonetheless only be put to use in entanglement-based QKD if chromatic dispersion effects of the links can be mitigated.
Without such mitigation, any reduction of $\sigma_\mathrm{J}$ is masked by chromatic dispersion effects. 
This becomes apparent when calculating $\sigma_{\mathrm{D}}$ for a fiber of length $L$, using the formula~\cite{SalehTeich}
\begin{equation}
\sigma_{\mathrm{D}} =\sigma_{\mathrm{\lambda}} D_{\mathrm{\lambda}} L
\label{eq:singlechannel}
\end{equation}
where $\sigma_\mathrm{\lambda}$ is the spectral width in wavelength of the propagating signal (typically $100$\,GHz) and $D_\mathrm{\lambda}$ is the wavelength-dependent dispersion coefficient.
$D_\mathrm{\lambda}$ takes positive (negative) values for anomalous (normal) dispersion, i.e. higher-energy photons traveling faster (slower).
Glass fibers and e.g. chirped fiber Bragg gratings can be manufactured to exhibit both positive and negative dispersion coefficients~\cite{Ramachandran2007}.
For a typical $D_\mathrm{\lambda}=+18$\,ps/nm/km at 1550\,nm, the dispersion amounts to $\sigma_\mathrm{D}\approx1400$\,ps for a $100$\,km inter-city link.
But even a comparably short $10$\,km link in an intra-city network such as in~\cite{Joshi2019, Dynes2019} exhibits about $140$\,ps of dispersion spread, which already poses a problem for high-end entanglement-based QKD implementations.
It is therefore of utmost importance for state-of-the-art QKD implementations to overcome dispersion effects.

Local dispersion compensation has been shown for prepare-and-send \cite{Boaron2018a} as well as entanglement-based \cite{Fasel2004, Aktas2016} QKD protocols by compensating right before or after the dispersive fiber channel. For entanglement-based protocols however, there is a unique method of dispersion compensation developed by Franson~\cite{Franson1992, Franson2009}.
He proposed to carry out so-called nonlocal dispersion compensation for entangled photon pairs by changing the dispersion in the transmission channel of one photon only.
In this way, we can exploit their wavelength correlations to restore temporal correlations which have been degraded due to chromatic dispersion effects.
In case of wavelength anti-correlation between entangled photons produced by spontaneous parametric down-conversion (SPDC), the total $\sigma_\mathrm{D}$ between Alice (A) and Bob (B) is equal to the sum of the channels' individual dispersion values:
\begin{eqnarray}
    \sigma_\mathrm{D}&=\sigma_\mathrm{D}^\mathrm{A}+\sigma_\mathrm{D}^\mathrm{B}\\
            &=\sigma_\mathrm{\lambda}(D_\mathrm{\lambda}^\mathrm{A}L^\mathrm{A}+D_\mathrm{\lambda}^\mathrm{B}L^\mathrm{B}).
 \end{eqnarray} 
Here, $\sigma_\mathrm{\lambda}$ is the photons' effective spectral width in wavelength, which is determined by the SPDC source and the filters in use.
Therefore, $\sigma_\mathrm{D}=0$ is possible for zero total dispersion ($D_\mathrm{\lambda}^\mathrm{A}L^\mathrm{A}=-D_\mathrm{\lambda}^\mathrm{B}L^\mathrm{B}$), reducing $\Delta T$ to contributions by $\sigma_\mathrm{C}$ and $\sigma_\mathrm{J}$ alone.

It is important to stress that the possibility of nonlocal dispersion compensation is a unique feature of entangled photon pairs.
The continuous cancellation of dispersion by acting on one photon only has no classical counterpart and can be considered a quantum advantage.
Also, it is an example of the versatility of entanglement-based QKD schemes.
The correlations of entangled photons, which naturally occur in SPDC in many different degrees of freedom, can be exploited in order to enhance secure key rates without compromising the degree of freedom used for actual key creation.
Entanglement can therefore be seen as a resource to further improve QKD protocols which are, in their basic form, concerned with one degree of freedom only.
In our case, this allows us to use just one device to compensate for dispersion in two different single-mode fibers carrying the entangled photons.
Thus, loss and additional complexity caused by the compensation device are the same as they would be in a single-channel experiment.

The principal feasibility of nonlocal dispersion compensation has been shown using a broad SPDC spectrum centered around the zero-dispersion wavelength of two similar fibers~\cite{Grieve2019}. Also, it was used to implement one measurement basis \cite{Lee2014} and to significantly violate Bell's inequality \cite{Zhong2013} in measurements on time-energy entangled photon pairs.
In this work, we develop the scheme further to demonstrate for the first time that nonlocal dispersion compensation can in fact be used for improving secure key rates in QKD applications.
We implement a full-fledged BBM92 QKD scheme based on polarization-entangled photon pairs sent along standard telecom fibers and show experimentally that by introducing negative dispersion in one channel only, tight timing correlations can be restored and key rates can be increased substantially.

\subsection{Experimental setup}
\begin{figure}
\centering\includegraphics[width=0.845\textwidth, right]{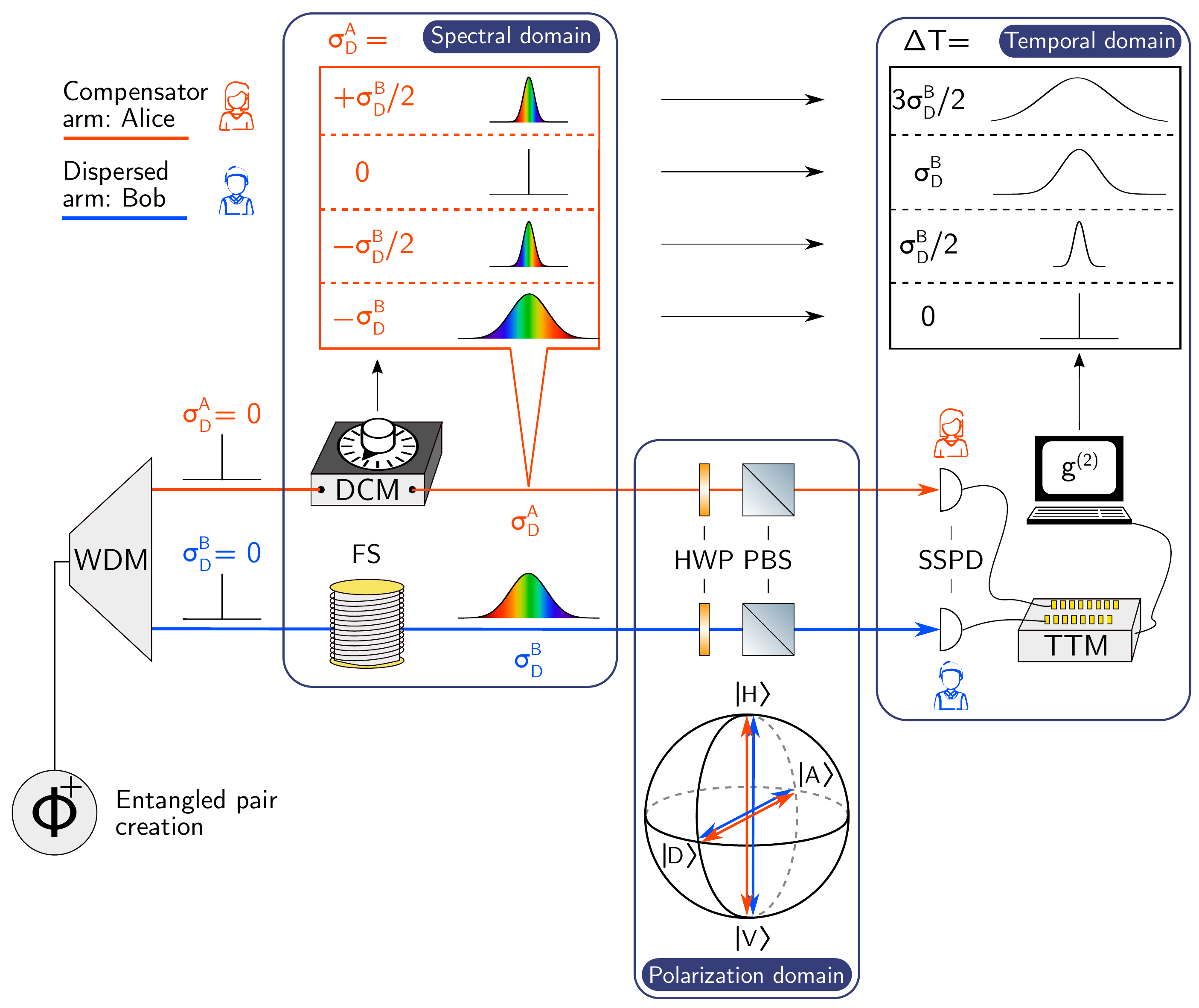}
\caption{Sketch of the experiment's working principle.
$\Phi^+$ denotes the source of photonic entanglement.
Two $200$\,GHz wavelength division multiplexing (WDM) channels guiding the entangled photon pairs are selected from the source spectrum.
The photon in Bob's arm passes a G.652 telecom fiber spool (FS) where it experiences positively signed dispersion $\sigma_\mathrm{D}^\mathrm{B}$.
The dispersion value in the other arm ($\sigma_\mathrm{D}^\mathrm{A}$) is manipulated via a dispersion compensation module (DCM).
Both photons are then detected by Alice and Bob, respectively.
They measure polarization by applying different settings of a half-wave plate (HWP) before a polarizing beam splitter (PBS).
The detection time is recorded using superconducting single-photon nanowire detectors (SSPD) connected to one time-tagging module (TTM).
We calculate the cross correlation between Alice's and Bob's time tags to generate the $g^{(2)}$ intensity-correlation functions displayed in \fref{fig:histograms}.
The resulting temporal width $\Delta T$ of the correlation function depends on the sum of the individual dispersion values $|\sigma_\mathrm{D}^\mathrm{A}+\sigma_\mathrm{D}^\mathrm{B}|$.
By introducing dispersion of equal magnitude and opposite sign, the initial non-dispersed timing correlations can be restored.
In reality, coherence time and detector jitter result in a minimum value for the $g^{(2)}$ spread, which is not visualized in the sketch for simplicity.}
\label{setup}
\end{figure}

The working principle of our experimental setup is shown in \fref{setup}. We use a Sagnac-type source of polarization-entangled photons with a type-0 phase-matched nonlinear crystal~\cite{Gayer2008}.
It is pumped with a continuous-wave laser at wavelength $\lambda_\mathrm{p}=775$\,nm, producing photon pairs with their spectrum centered at approximately $1550$\,nm via SPDC.
The down-converted photons are coupled into a single-mode fiber (SMF).
Obeying energy conservation, $\lambda_\mathrm{p}\approx(\lambda_\mathrm{s}+\lambda_\mathrm{i})/4$ holds for individual entangled photon-pairs, where s (i) denotes signal (idler) photons.
This relation can be used to select entangled signal and idler photons from the full spectrum.
To this end, we use dense wavelength division multiplexing (DWDM) top-hat filters with $200$\,GHz broad spectral transmission~\cite{ITUG694, Grobe2008}.
With two such filters, we realize wavelength channels centered at $1549.32$ and $1550.92$\,nm, respectively, each carrying one of the entangled photons.
We align the source such that the photons in these respective color channels are maximally entangled in their polarization degree of freedom, forming the Bell state
\begin{eqnarray}
    |\phi^{+}\rangle_{\mathrm{pol}}&=1/\sqrt{2}(|H_\mathrm{s},H_\mathrm{i}\rangle+|V_\mathrm{s},V_\mathrm{i}\rangle)\nonumber\\
     &=1/\sqrt{2}(|D_\mathrm{s},D_\mathrm{i}\rangle+|A_\mathrm{s},A_\mathrm{i}\rangle)
\label{eq:bellstate}
\end{eqnarray}
where $H$ ($V$, $D$, $A$) denotes horizontal (vertical, diagonal, antidiagonal) polarization.

The signal photons are injected into a 6.46\,km long G.652 telecom fiber with $D_\lambda=+16.7\pm1.0$\,ps/nm/km as specified by the manufacturer~\cite{Fionec, CorningSMF28}, resulting in a calculated total dispersion of $\sigma_D^B/\sigma_\lambda=+107.9\pm6.5$\,ps/nm.
In order to nonlocally compensate for this dispersion in Bob's arm, Alice’s channel carrying the idler photons is connected to a Teraxion Clearspectrum T2506 dispersion compensation module (DCM).
According to the display's reading, it can introduce dispersion values $\sigma_D^A/\sigma_\lambda$ ranging from $-170$ to $+170$\,ps/nm in $10$\,ps/nm steps~\cite{DCMspecs}.

We simulate a long-distance scenario of $300$\,km distance in terms of loss by introducing attenuation of about $30$\,dB in each channel.
This is done by decreasing the SMF coupling efficiency at the detectors. 
We calculate the loss via the Klyshko or heralding efficiency, i.e. by determining the ratio between correlated photons and all detector clicks (less noise counts)~\cite{Klyshko1980}.

Alice and Bob determine the polarization state of their respective photons using polarization analysis modules, consisting of a half-wave plate (HWP) and a polarizing beam splitter (PBS), and detect the photons via superconducting single-photon nanowire detectors (SSPD) of the Single Quantum Eos series connected to the same time tagging module (TTM) Ultra 8 by Swabian Instruments.
We report a constant background noise level of $160$\,kcps (Alice) and $175$\,kcps (Bob).

\section{Results}

In order to quantify the QKD protocol's performance, Alice and Bob record a time tag and the HWP’s angle setting for each of their measurement events.
The HWP settings correspond to different polarization measurements ($0^\circ=H$, $22.5^\circ=D$, $45^\circ=V$, $67.5^\circ=A$), where equal (orthogonal) settings at Alice and Bob correspond to correct (erroneous) results, owing to the desired Bell state described in eq. \eref{eq:bellstate}.
These measurements are carried out for different settings of the DCM.

\begin{figure}[htpb!]
   \centering\includegraphics[width=0.845\textwidth, right]{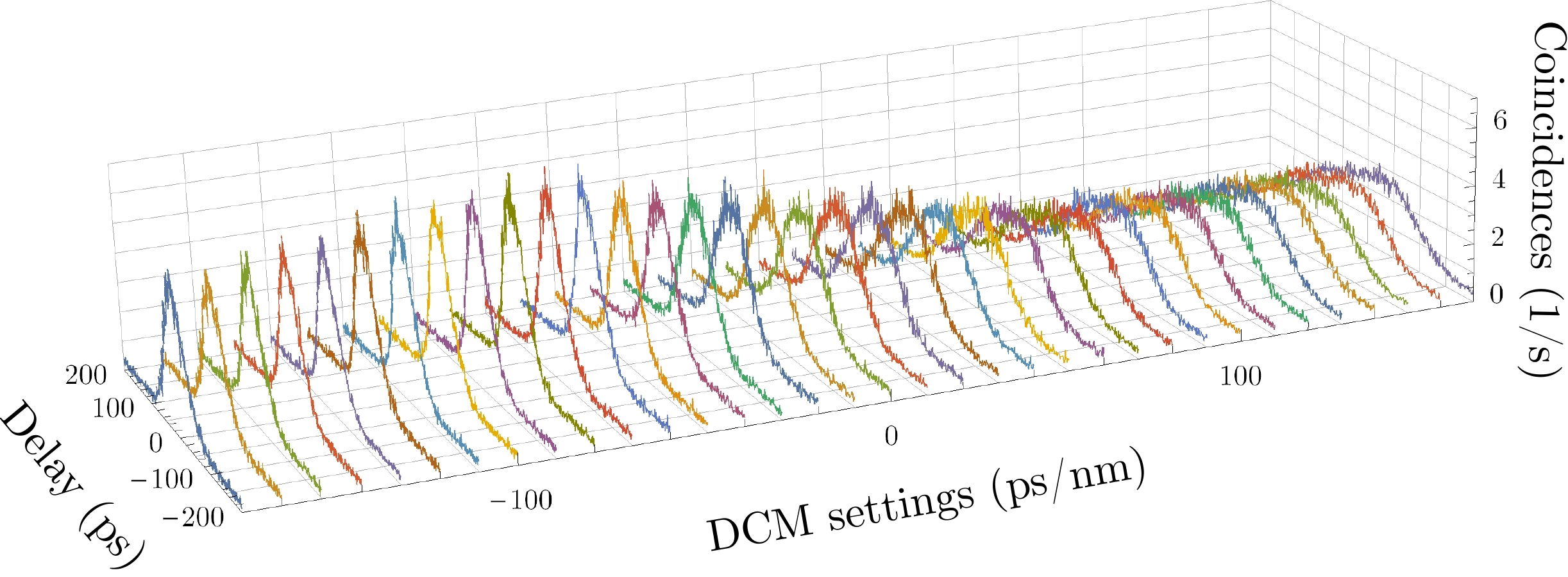}
  \caption{Histograms of coincident events according to $HH$ polarization measurements per relative delay between Alice and Bob in $1$\,ps bins, normalized to $1$\,s. $VV$, $DD$ and $AA$ (not depicted) show similar behavior. Each histogram corresponds to a different setting of the dispersion compensation module (DCM). Note that DCM settings $>0$\,ps/nm are equivalent to simulating another dispersive fiber in Alice's arm. The peak values decrease for broader temporal spreads due to constant total coincidence rates.}
  \label{fig:histograms}
\end{figure}

$\Delta T$ can then be determined for each DCM setting by plotting a histogram showing the number of time tags at Alice and Bob per temporal delay between them (for $HH$ see \fref{fig:histograms}).
The full width at half-maximum (FWHM) of these histograms corresponds to $\Delta T$.
We clearly observe dispersion-induced changes of $\Delta T$ when tuning the DCM settings over their full range of $-170$ to $+170$\,ps/nm.
The positively signed temporal dispersion $\sigma_\mathrm{D}^B$ of the fiber can be counteracted by introducing negative dispersion coefficients in the DCM, reducing $\Delta T$ to contributions from detector jitter and coherence time only.
Setting the DCM to positive dispersion values however simulates a long-distance fiber link in Alice's channel, thus further increasing $\Delta T$.

\begin{figure}
\centering\includegraphics[width=0.845\textwidth, right]{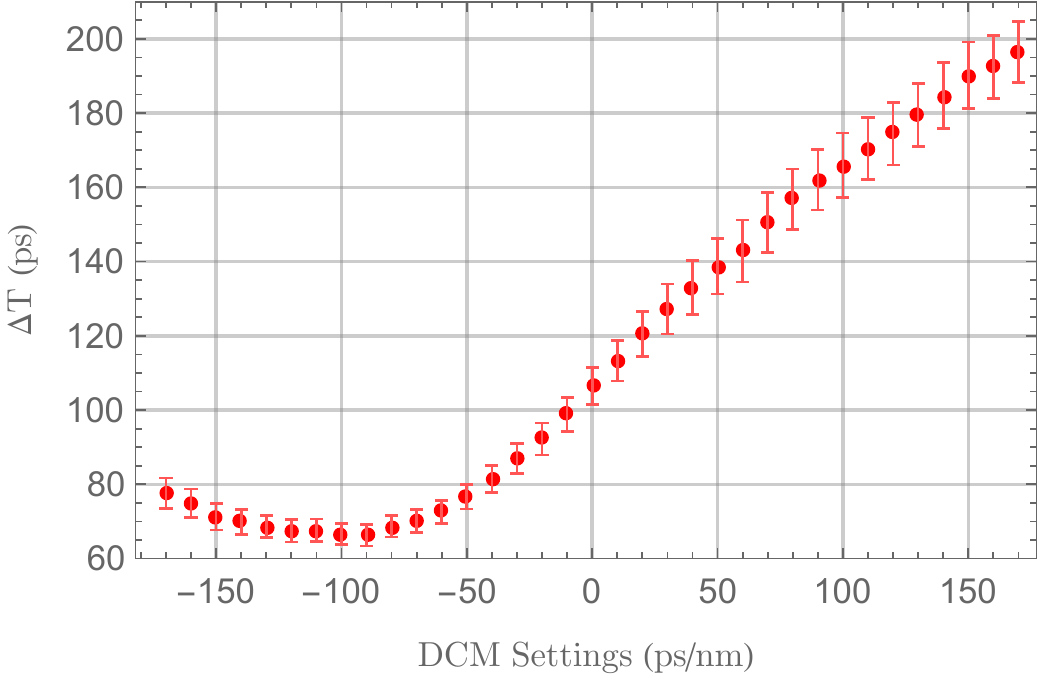}
\caption{Average temporal dispersion $\Delta T$ as a function of the dispersion compensator module (DCM) settings. The values of $\Delta T$ were acquired by fitting Gaussian functions to all histograms. Each data point represents the average FWHM of all four equal polarizer settings. Errors bars are plotted assuming Poissonian errors of the detection rates. Optimal dispersion compensation was found for the DCM setting of $-90$\,ps/nm, where $\Delta T$ is limited by the detector jitter $\sigma_\mathrm{J}=66\pm3$\,ps. For DCM values of $-100$\,ps/nm and lower, the dispersion in Bob's arm is being overcompensated, thus again increasing the total temporal uncertainty $\Delta T$.}
\label{fig:FWHMs}
\end{figure}

\Fref{fig:FWHMs} shows the average FWHMs of all correct correlations for each of these DCM settings, which we extracted from Gaussian fits to the experimental data.
The minimal value of $\Delta T=66\pm3$\,ps was found for a DCM display reading of $-90$\,ps/nm.
The mismatch with the calculated location of the minimal value at $-107.9$\,ps/nm, which amounts to a $5\%$ deviation in the considered range, can be explained by DCM imperfections and/or deviations of the fiber's dispersion specification.

Summarizing, we were able to tune $\Delta T$ from $66$ to $197$\,ps with our method, inducing both normal and anomalous dispersion in Alice's arm.
Such dispersion manipulation of entangled photon pairs must be understood as a nonlocal process~\cite{Franson2009}.
This is because the dispersion we introduced in Bob's arm was compensated for by changing solely the dispersion in Alice's arm, while Bob's dispersion value stayed constant.

\section{Discussion}

\begin{figure}
    \centering\includegraphics[width=0.845\textwidth, right]{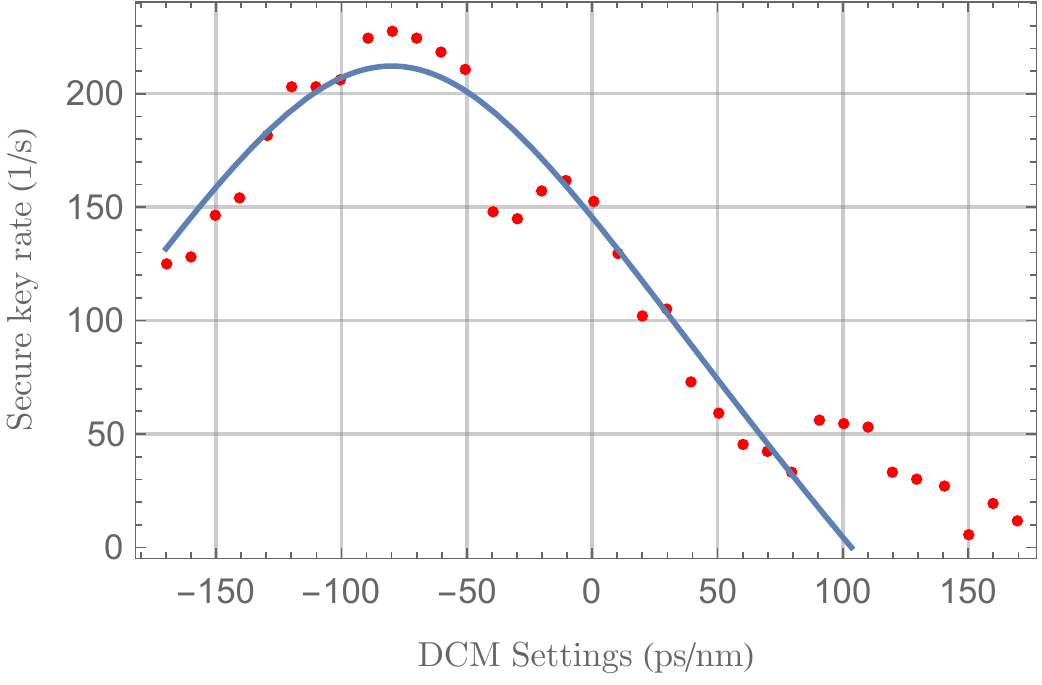}
    \caption{Secure key rates for individually optimized coincidence windows $t_\mathrm{CC}$ vs. dispersion compensation module (DCM) settings. For a DCM setting of $-80$\,ps/nm, we observed a secure key rate of $228.3$\,bits/s as compared to just $6.1$\,bits/s for the least favorable setting. Thus, we observe a 37-fold improvement of the key rate by nonlocal dispersion compensation. The blue graph shows our simple model (see \ref{chap:model}) which captures the main behavior of the experimental data.}
    \label{fig:key}
\end{figure}

We will now investigate the secure key rate implications of these dispersion-induced changes to $\Delta T$.
For the above histograms, only correct polarization correlations were used to determine $\Delta T$, since the erroneous ones mainly consist of the noise floor due to high fidelity of our entangled state.
For carrying out the actual QKD protocol however, all correlations have to be used, since Alice and Bob must not publicly communicate their polarization measurement outcomes, but only their time tags and basis choices.
Once they have done so, they have to agree on a delay including a tolerance interval, the so-called ``coincidence window'' $t_\mathrm{CC}$, to define those events that are used for key creation (``coincidences'').
Naturally, the optimal choice of $t_\mathrm{CC}$ in terms of secure key rate strongly depends on $\Delta T$ of the acquired histograms:
if the histogram is flattened due to dispersion, one is forced to use a larger $t_\mathrm{CC}$ in order to collect as many coincidences as possible.
The number of erroneous detection events however increases proportionally to $t_\mathrm{CC}$, therefore inflating the quantum bit error rate.
This explains the behavior of the secure key rate as shown in \fref{fig:key}.
Here, we calculated the maximal secure key rate for every setting of the DCM module as the average of horizontal-vertical and diagonal-antidiagonal basis settings.
We numerically optimized $t_\mathrm{CC}$ to acquire the highest possible secure key rate for each DCM setting individually (see \ref{chap:seckey}).
The overall maximal secure key rate of $228.3$ bits/s is found for the DCM setting at $-80$\,ps/nm.
Compared to the lowest acquired value of $6.1$\,bits/s, our nonlocal dispersion compensation scheme therefore resulted in a $37$-fold increase in secure key rate. 

This demonstrates the detrimental effect of dispersion and its overcoming by nonlocal dispersion compensation.
The dispersion-induced degradation of our QKD system's error rate could be annihilated with our method.
Furthermore, the observed overall behavior is in good agreement with our theoretical model as can be seen in \fref{fig:key}.
Also note that conventional compensation of dispersion in both channels would require the use of \emph{two} lossy DCM modules. Assuming a hypothetical second module with the same attenuation of $4.56$\,dB, calculations using our model show that the maximum obtainable key rate would only have been $38.9$\,bits/s in such a \emph{local} dispersion compensation case.
Details on the model are provided in \ref{chap:model}.

We have shown that chromatic dispersion acting on an entangled photon pair can be compensated in a nonlocal manner by manipulating only the dispersion experienced by one of the two entangled photons.
Doing so, the tight original timing correlations of the entangled source's emission process can be retrieved by exploiting their non-degraded wavelength anticorrelations.
To the best of our knowledge, this is the first experimental demonstration of improving secure key rates in QKD via nonlocal dispersion compensation.

\section{Conclusion}
We have devised a QKD implementation over a 6.46\,km fiber link and successfully managed to increase the resulting secure key rates by compensating for chromatic dispersion in a nonlocal manner.
Utilizing wavelength anticorrelations of polarization-entangled photons to counteract temporal broadening, we have shown a 37-fold gain of key rates compared to the least favorable dispersion configuration.
For this nonlocal compensation scheme, a ready-to-use off-the-shelf DCM and patch fibers were deployed, with no need for further alignment of sensitive components.
Additionally, our experiment was designed to match real-life loss scenarios.
Since one device is enough to compensate a two-channel QKD scheme, the DCM insertion loss is the same as it would be in a single-channel experiment.
The scheme is therefore ideally suited for real-world applications.

Our findings can easily be generalized to substantially longer fiber links, where control of dispersion is a prerequisite for obtaining high key rates or even any key at all.
Taking a 400\,km link as an example, the secure key rate can be increased by a factor of 400 to about 10\,bits/s with our method.
This estimate is ignoring noise counts, e.g., from parallel classical traffic, which would increase the necessity of tight timing correlations even further.
Since excellent timing precision is obligatory for state-of-the-art QKD, our scheme can help to enhance any in-fiber entanglement-based QKD system. It is straightforward to adapt it to high-dimensional entanglement or other degrees of freedom, e.g. time-energy entanglement.
Concluding, we are convinced that the nonlocal dispersion compensation scheme presented in this work will be an essential component for future implementations of fiber-based QKD networks.

\ack
We acknowledge European Union’s Horizon 2020 programme grant agreement No. 857156 (OpenQKD) and the Austrian Academy of Sciences. We also want to thank Josef Vojtech of CESNET (Prague) for providing us with dispersion compensation equipment, Siddarth Koduru Joshi of NSQI (Bristol) for lending us a DWDM device and S\"oren Wengerowsky of IQOQI (Vienna) for fruitful discussions.

\appendix
\section*{Appendix}
\setcounter{section}{1}
\subsection{Secure key rate calculation\label{chap:seckey}}
To quantify the improvements achieved by our dispersion compensation scheme, we calculated the secure key rate in the asymptotic limit of infinite key size for each DCM setting.
In order to do so, one first needs to calculate the quantum bit error rate (QBER, $E$)~\cite{Xu2020}, which is defined as the ratio of erroneous coincidences to total coincidences. It can be written as
 \begin{equation}
    E=\frac{CC_\mathrm{err}}{CC_\mathrm{corr}+CC_\mathrm{err}}
\label{eq:qberexp}
\end{equation}
where $CC_\mathrm{err}$ ($CC_\mathrm{corr}$) is the number of erroneous (correct) coincidences per second.
Using $E$, one can calculate the lower bound for the secure key rate $R_s$ in the infinite-key limit is using the formula~\cite{Ma2007}
\begin{equation}
    R_\mathrm{s}=CC_\mathrm{tot}\cdot\big(1-(1+f)H_2(E)\big).
    \label{eq:seckey}
\end{equation}
Here, $CC_\mathrm{tot}=CC_\mathrm{corr}+CC_\mathrm{err}$ is the total number of coincidences per second, $f=1.1$~\cite{Elkouss2009} is the bi-directional error correction efficiency and $H_2(x)$ is the binary entropy function~\cite{Ma2007}.
$CC_\mathrm{corr}$ and $CC_\mathrm{err}$ both depend on the chosen coincidence window $t_\mathrm{CC}$, which has been determined numerically to optimize $R_\mathrm{s}$ for each DCM setting.

\subsection{Fitting model\label{chap:model}}
The blue curve in \fref{fig:key} represents our model of the secure key rate behavior depending on the DCM settings. For this model, $CC_\mathrm{tot}$ in eq. \eref{eq:seckey} is calculated using the source brightness $B$, channel losses $\eta_i$, and a correction factor $s$ (ignoring noise counts):
\begin{equation}
CC_\mathrm{tot}=s B\eta_A\eta_B.
\label{eq:CC}
\end{equation}
The QBER $E$ in eq. \eref{eq:qberexp} and eq. \eref{eq:seckey} is modeled as
\begin{equation}
    E=
     \frac{s B\eta_A\eta_Be_{\mathrm{o}}+\xi/2}{s B\eta_A\eta_B+\xi}
    \label{eq:qber}
\end{equation}
where $e_{\mathrm{o}}$ is the probability of erroneous detection due to optical imperfections of source and polarization analyzers, $DC_i$ are the noise counts per detector and
\begin{equation}
    \xi=(B\eta_A+2DC_A)(B\eta_B+2DC_B)\sqrt{\sigma_\mathrm{C}^2+\sigma_\mathrm{J}^2+\sigma_\mathrm{D}^2}
    \label{eq:acc}
\end{equation}
is the rate of coincident counts which arise by chance due to the finite coincidence window $t_\mathrm{CC}$ and not due to an actual photon pair.
$\xi$ is divided by $2$ in the numerator of eq. \eref{eq:qber} because only half of these ``accidental'' clicks contribute to erroneous coincidences with orthogonal polarizer settings and the other half is registered as correct.
For simplicity, we do not account for the numerical optimization of the coincidence window in our model, but set $t_\mathrm{CC}=\Delta T$.
The factor $s=\mathrm{erf}[\sqrt{\mathrm{ln}(2)}]=0.76$ in eq. \eref{eq:CC} and eq. \eref{eq:qber} accounts for the fact that true coincidence clicks originating from photon pairs follow a Gaussian distribution with FWHM $\Delta T$, i.e. clicks outside the coincidence window at the ``tails'' of the distribution are lost.

\Fref{fig:key} shows this key rate model for the following parameters in use: $B=5.75\times10^8$\,cps, $\eta_1=29.05$\,dB, $\eta_2=29.31$\,dB, $DC_1=1.4\times10^5$\,cps, $DC_2=1.75\times10^5$\,cps, $e_\mathrm{o}=0.01$,  $\sigma_\mathrm{J}=66$\,ps, $\sigma_\mathrm{C}=0$\,ps.
All modeling parameters were estimated from experimental data.
The model was offset by $-80$\,ps in order to fit the data, although we expected optimal compensation to take place at $-90$\,ps.
This discrepancy is most likely due to statistical fluctuations in the measured count rates and will be subject of future studies.
Further deviations between our model and the observed $R_\mathrm{s}$ can be explained by the fact that our simple model does not capture the numerically obtained optimal coincidence windows, which is especially important for low key rates, and that all underlying distributions were assumed to be perfectly Gaussian for simplicity.
Nevertheless, we observe that our model correctly captures the main features of the experimentally obtained key rates and thus explains the functional dependence of secure key rate on nonlocal dispersion compensation.

\subsection{Improvements using non-local dispersion compensation over long distances\label{chap:limits}}
\begin{figure}[htbp!]
    \centering\includegraphics[width=0.7\textwidth]{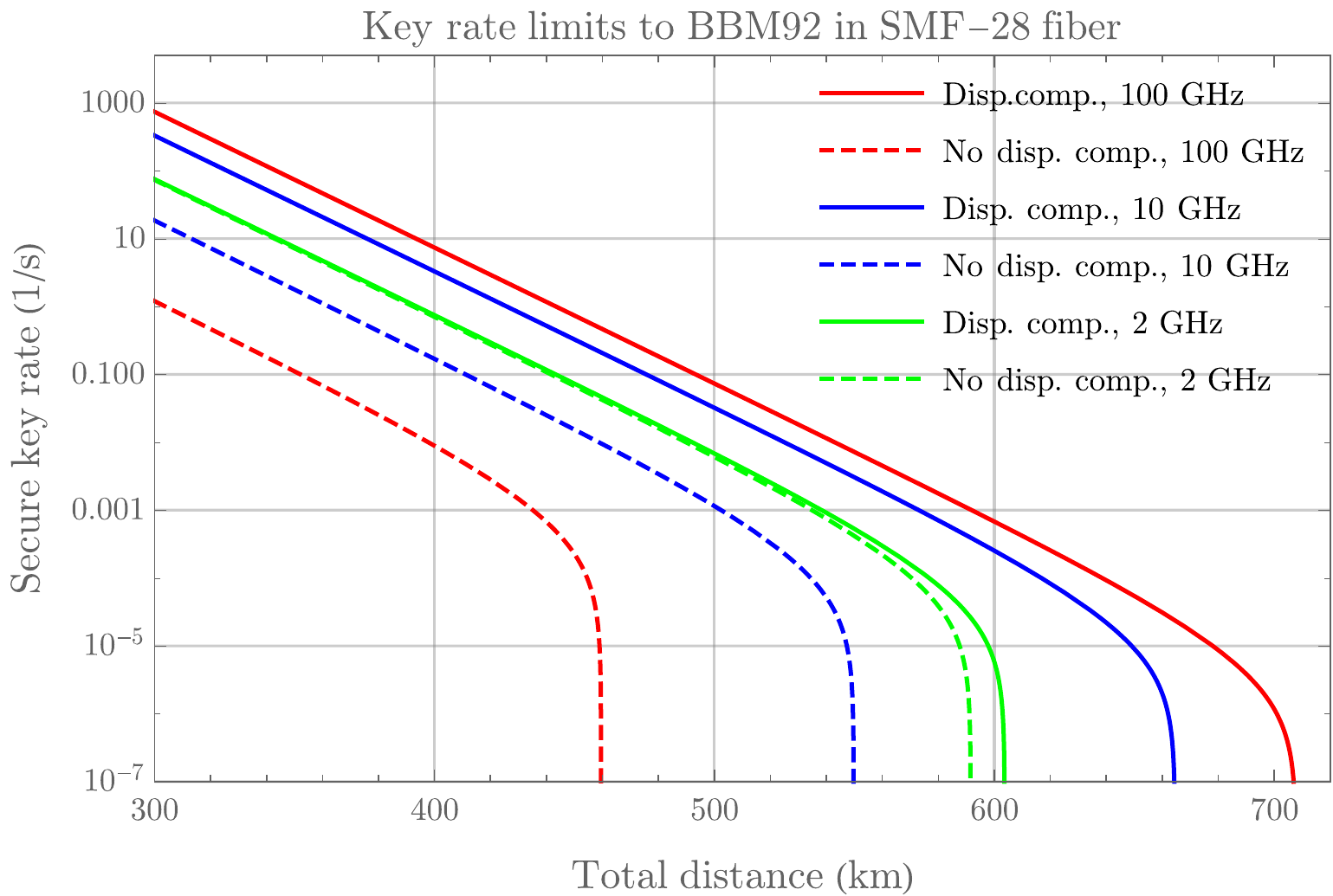}
    \caption{Expected secure key rates over total communication distance in kilometers of SMF-28 fiber, assuming a BBM92 protocol with symmetric channel loss and chromatic dispersion. For each distance and spectral width, the source brightness is optimized. Solid (dashed) lines refer to QKD systems with (without) dispersion compensation. The different colors refer to the width of the selected entangled photon's joint spectral amplitude centered at 1550\,nm. The behavior clearly shows that broader spectra using nonlocal dispersion compensation are favorable as compared to narrow spectra with low total dispersion.}
    \label{fig:limits}
\end{figure}

Using the model introduced above, we estimate the maximum achievable distance and key rate for a fiber-based BBM92 system along standard SMF-28 fiber. Fig. \ref{fig:limits} shows calculations for identical fiber channels from source to Alice and Bob, with attenuation of $0.2$\,dB/km and $+18$\,ps/km/nm chromatic dispersion. We assume state-of-the-art parameters for all curves: dark counts $DC=100$\,cps per detector, $1$\,\% optical error $e_\mathrm{o}$ and $20$\,ps detector jitter $\sigma_\mathrm{J}$. Brightness values $B$ are optimized for each curve individually, with values ranging from $6.6\times 10^6$ to $2.5\times10^9$\,cps.
Different spectral widths of the photons lead to different efficiencies of our non-local dispersion compensation scheme. If the photons are narrowly filtered, dispersion contributes much less to $\Delta T$ than coherence time (see eq. \eref{eq:deltaT}). Thus, in the case of an optimal time-bandwidth product at 2\,GHz, dispersion compensation can only marginally increase key rate and  maximal communication distance. The broader the spectrum however, the more key rate and maximum distance can be gained with our scheme, and the 2\,GHz case can be outperformed: for 10\,GHz (100\,GHz), the maximum distance is increased by 115\,km (250\,km). Also key rates can be increased substantially. Optimizing for 300\,km distance, dispersion compensation allows for a gain in key rate from 5 to 1036\,bits/s in the 100\,GHz case and from 49 to 438\,bits/s with 10\,GHz. Broader photon spectra allow for a higher maximum key rate when using dispersion compensation, since one is not limited by long coherence times. In addition to enhanced performance, using broad spectra also has the advantage of substantially less complex source designs. This is because sources of photonic entanglement that exhibit both narrow photon spectra and high brightness require the use of cavities and/or waveguides \cite{Anwar2020arx}.

\section*{References}
\bibliographystyle{unsrt}
\bibliography{dispersionlibrary}

\begin{thebibliography}{10}
\bibitem{DCMspecs}
Unfortunately, we cannot give more detailed information about the device, since
  production is discontinued by the manufacturer and neither data sheet nor
  manual are available to us.
\end{thebibliography}

\section*{Author Contributions}
S.N. and R.U. devised the experiment. S.N. and D.R. built the measurement setup. D.R. collected measurement data. S.N., D.R. and M.B. analyzed the data. R.U. supervised the work. S.N., M.B. and R.U. contributed to write the paper.

\section*{Competing Interests}
The authors declare no competing interests.

\section*{Additional Information}
\textbf{Correspondence} and requests for materials should be addressed to S.N. and R.U.

\end{document}